 \documentclass[preprint2]{aastex}

\usepackage{graphicx}
\def\dsm{$M_\odot$}
\def\dsr{$R_\odot$}

\shorttitle{The Effects of Binaries on the CMDs of Young-Age Star Clusters}
\shortauthors{Wuming Yang}

\begin{document}


\title{The Effects of Binary Stars on the Color-Magnitude Diagrams of Young-Age Massive Star Clusters}
\author{Wuming Yang}
%\author{Wuming Yang\altaffilmark{1} }
%\affil{$^{1}$Department of Astronomy, Beijing Normal University,Beijing 100875, China}
\affil{Department of Astronomy, Beijing Normal University,Beijing 100875, China}
\email{yangwuming@bnu.edu.cn, yangwuming@ynao.ac.cn}

\begin{abstract}
Extended main-sequence turnoffs (eMSTO) have been observed in the 
color-magnitude diagram (CMD) of intermediate-age and young star clusters. 
The origin of the eMSTO phenomenon is still highly debated. Calculations
show that the blue and faint (BF) stars in the CMD of NGC 1866 are hydrogen
main sequence (MS) + naked He star systems. The He star derives from the 
massive star of a binary system. The BF stars and the red and faint MSTO 
stars belong to the same stellar population. The values of $m_{F336W}$ and
$m_{F336W}-m_{F814W}$ of the BF stars are mainly determined by the masses of He 
stars and H-MS stars, respectively. The behaviors of the BF stars 
in the CMD are well explained by the H-MS + He-star systems. The BF 
stars provide a strict restriction on the age of the stellar population. 
Moreover, the bimodal MS of NGC 1866 can also be reproduced by a younger 
binary population. The calculations show that part of the blue and bright 
(BB) MS stars of NGC 1866 are H-MS + He-star systems, H-MS + white dwarf 
systems, and merged stars in a binary scenario. The H-MS stars of 
the H-MS + He-star systems for the BB stars are significantly more massive than 
those of the BF stars. Once the H-MS + He-star systems and their membership
in NGC 1866 are confirmed, the extended star-formation histories and the effects 
of binaries can be confirmed in the young star cluster.
\end{abstract}


\section{INTRODUCTION}
In the classical theory of star formation, a star cluster is considered to be
composed of stars belonging to a simple, single stellar population (SSP)
with a uniform age and chemical composition. However, the discoveries of 
double or extended main-sequence turnoffs (eMSTO) \citep{mack07, glat08,
gira09, goud09, milo09} in the color-magnitude diagram (CMD) of 
intermediate-age star clusters in the Large Magellanic Clouds (LMC) 
are challenging the classical hypothesis. The phenomenon of eMSTOs
has been interpreted as meaning that the clusters have
experienced extended star-formation histories (eSFH) with
durations of $\sim$100-700 Myr \citep{glat08, mack08, milo09, gira09, rube10, 
goud11, goud14, corr14}. As an alternative interpretation, eMSTOs are thought to be 
due to coeval populations with different rotation 
rates \citep{bast09, yang13, lic14, bran15, dant15, nied15a} or due to interacting
binaries \citep{yang11, liz12, liz15}. Moreover, stellar variability \citep{sali16, degr17}
and a variable overshoot of the convective core of stars \citep{yang17} may play a
potential role in shaping the eMSTO regions as well. The nature of the eMSTO phenomenon 
is still highly debated.

The problem of the origin of eMSTOs becomes more complicated when they
and split main sequences (MS) are found in the CMD of young clusters (less than
$\sim 400$ Myr) in the LMC. The split MS was first discovered in the CMD 
of young cluster NGC 1844 \citep{milo13}. However, the eMSTOs of young clusters
were first observed in NGC 1856 by \cite{corr15} and \cite{milo15}. Young
cluster NGC 1856 hosts an eMSTO and a double MS, which changes our understanding 
of young clusters in the LMC. The eMSTO of NGC 1856 can be interpreted as the superposition
of two main populations having the same age but different rotation rates \citep{dant15}
or as the effects of a variable overshoot of the convective core of stars \citep{yang17}.
\cite{bast17} have inferred the existence of rapidly rotating stars
in NGC 1856 and NGC 1850 from H-alpha excess fluxes likely being due to 
so-called Be candidate stars.

Moreover, eMSTOs and bimodal MS have been observed in young clusters 
NGC 1755 \citep{milo16}, NGC 1850 \citep{bast16, corr17}, and 
NGC 1866 \citep{milo17}. Neither stellar populations with different
ages only, nor coeval stellar populations featuring a distribution of 
stellar rotation rates, properly reproduce the observed split MS and 
eMSTO \citep{milo17, corr17}. eMSTOs were found in young
clusters NGC 330, NGC 1805, NGC 1818, and NGC 2164 as well \citep{lic17}.
\cite{lic17} also show that the observed eMSTOs cannot be explained by
stellar rotation alone. Similar cases have been found in intermediate-age 
star clusters NGC 1987 and NGC 2249, whose eMSTOs cannot be explained
solely by a distribution of stellar rotation rates \citep{goud17}.
For these clusters, a combination of rotation and an age spread
seems to be required to explain observational 
results \citep{milo16, milo17, corr17, lic17, goud17}.

\cite{piat17} analyzed the data of young cluster NGC 1971 and found 
that NGC 1971 exhibits an eMSTO originated mostly by a real age spread.
Moreover, \cite{dupr17} obtained the spectra of 29 eMSTO stars in NGC 1866.
The direct spectroscopic measures clearly demonstrate the presence of 
rapidly rotating stars that are cooler than a population of slowly rotating 
objects, arguing for an actual spread in age of NGC 1866. However,
\cite{lema17} studied the chemical composition of several Cepheids
located in NGC 1866 and found that six Cepheids have a homogeneous
chemical composition and are consistent with the red giant branch in the cluster.
Their analysis shows that the Cepheids belong to the same stellar population.
In line with the comment of \cite{milo17} on their observational results of NGC 1866, 
these observations raise many more questions than they solve.

\begin{figure}
\centering
\includegraphics[scale=0.7, angle=-90]{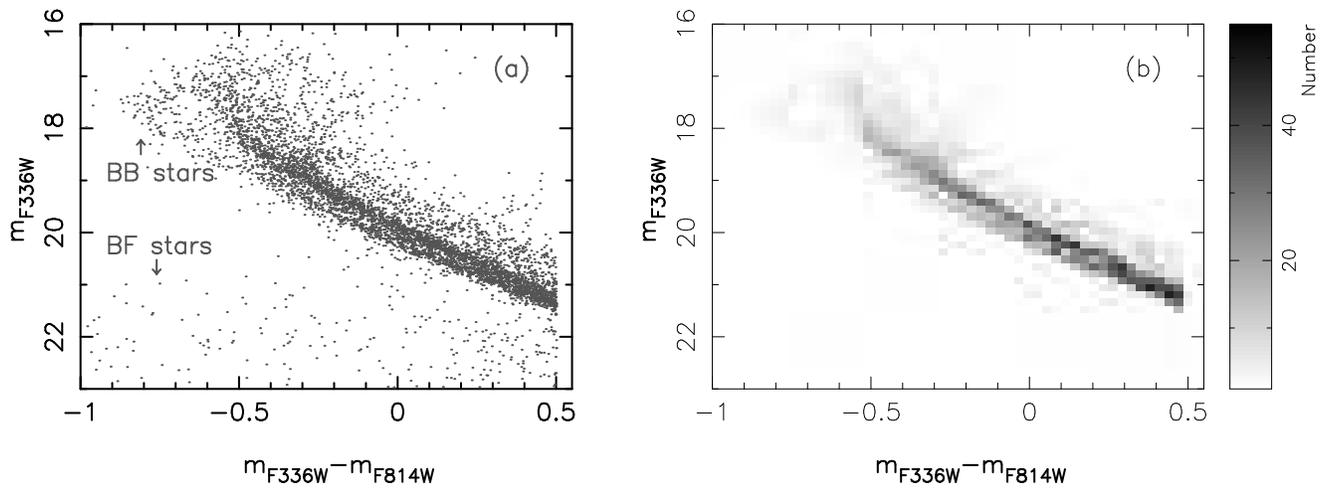}
\caption{The color-magnitude diagrams of NGC 1866 obtained by \cite{milo17}.
The blue and bright stars on the top left of panel $a$ are labelled as BB stars,
while the blue and faint stars on the bottom left of panel $a$ are labelled as 
BF stars. The grey scale in panel $b$ is proportional to the number of stars.}
\label{fig1}
\end{figure}

In addition to the main characteristics of the eMSTO and bimodal MS described 
by \cite{milo17}, Figure \ref{fig1} shows that NGC 1866 has two main
MSTOs and that there are many blue and bright (BB) stars and blue and 
faint (BF) stars in the CMD of NGC 1866. 
There is a gap between the BB stars and the blue or red MS stars
and an upper limit of luminosity for the BF stars; that is, 
the value of $m_{F336W}$ for most of the BF stars is larger than 
$\sim 21$. An age spread or the effects of rotation cannot 
produce such BF stars. The behaviors of the BF and 
BB stars in the CMD might result from binaries.
If the BF and BB stars are members of NGC 1866,
they provide different perspectives on the 
populations of NGC 1866 and aid us in understanding the 
nature of the eMSTO phenomenon in young clusters.

In the present work, we mainly focus on whether the characteristics of NGC 1866
can be reproduced by binaries. The paper is organized as follows. Some 
initial assumptions are given in Section 2, calculation results are
shown in Section 3, and the results are discussed and summarized in Section 4.

%__________________________________________________________________
\section{STELLAR MODELS AND POPULATION SYNTHESIS}
For a binary system, the mass of the primary star, $M_{1}$, is generated in terms of
the lognormal initial mass function (IMF) of \cite{chab01}.
The mass of the secondary star is then determined by $qM_{1}$,
where the $q$ is the ratio of the mass of the secondary to that 
of the primary and is generated according to an assumed 
distribution. The distribution of separations ($a$) between the primary 
and the secondary stars is assumed to be constant in $\log a$ \citep{han95}:
\begin{equation}
 an(a)=\alpha_{a},
\end{equation}
where $\alpha_{a} \approx 0.12328$. The eccentricity ($e$) of each
binary system is assumed to be a uniform distribution within
$0-1$. 

In this study, the initial metallicity $Z$ was fixed at $0.008$. Once the initial
distributions of the masses $M_{1}$ and $M_{2}$ ($qM_{1}$), separation $a$,
and eccentricity $e$ were given by Monte Carlo simulation, the sample 
was evolved to a given age by using the Hurley rapid single/binary evolution 
codes \citep{hurl00, hurl02} to obtain the luminosities and effective 
temperatures of the stellar population with the given age. 

The metallicity $Z$ was converted into [Fe/H] by [Fe/H] 
$= \lg(Z/X)- \lg(Z/X)_{\odot} \simeq \lg(Z/Z_{\odot})$. There is
a bitter controversy between helioseismology and observation about 
solar metallicity \citep[and the references therein]{yang16}, but the value of
$Z_{\odot}$ in our calculations is $0.02$. The quantities ([Fe/H], 
$T_{\rm eff}$, $\log g$, $\log L$) of evolutionary models were then transformed 
into colors and magnitudes using the color transformation tables of \cite{leje98}.
In computing their colors and magnitudes, the binaries with 
$a \le 10^{4} R_{\odot}$ were treated as a single point-source object 
according to the formulas in \cite{zhan04}.
A distance modulus $(m-M)_{0}=18.3$ and reddening $E(B-V)=-0.01$
were adopted in our calculations.

\section{CALCULATION RESULTS}
\subsection{Results of a Uniform $q$ Distribution}
\begin{figure}
 \includegraphics[scale=0.6, angle=-90]{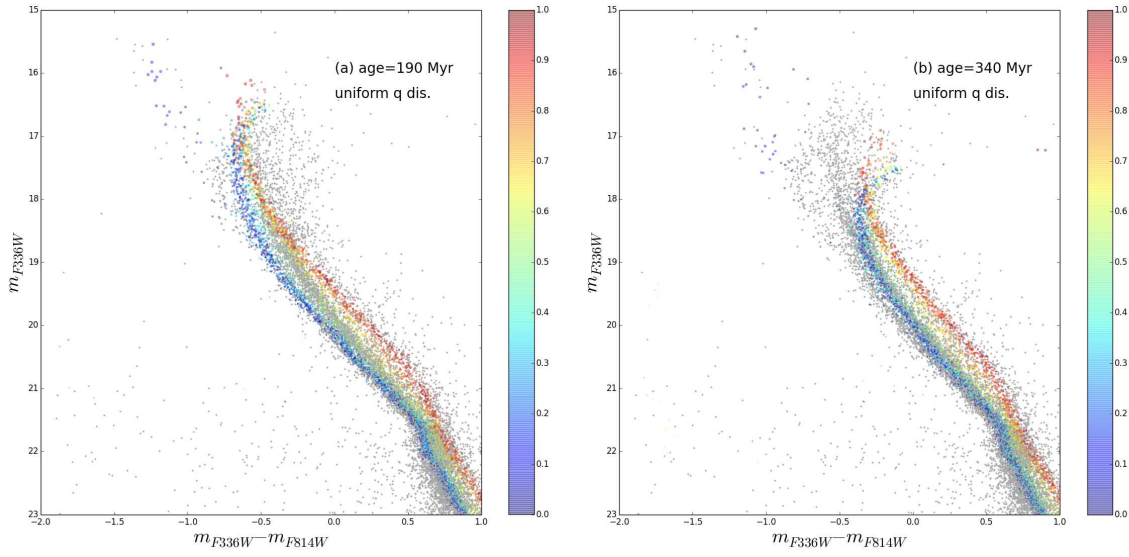}
\caption{CMDs of the observed and simulated stellar populations.
The grey dots refer to the observed data of NGC 1866 \citep{milo17}.
The color dots represent simulated stellar populations.
The size and the color of the color dots are proportional to stellar mass
and mass-ratio $q$, respectively. The value of $q$ of merged stars is $0$.}
\label{fig2}
\end{figure}

There are about $9900$ objects with $15.0\leq m_{F336W}\leq 23.0$ and 
$-2.0\leq m_{F336W}-m_{F814W}\leq 1.0$ in the observed data of NGC 1866 \citep{milo17}.
With the assumption that the mass ratio $q$ is a uniform distribution within
$0-1$, a sample was evolved to given ages. The CMDs of the simulated 
populations with different ages are shown in Figure \ref{fig2}.
In each panel of Figure \ref{fig2}, there are about $4000$ simulated objects 
with $15.0\leq m_{F336W}\leq 23.0$ and $-2.0\leq m_{F336W}-m_{F814W}\leq 1.0$, 
where the merged stars make up about $13\%$. Others are binaries. 
In our synthesized populations, we included observational errors taken
to be a Gaussian distribution with a mean value of $0$ and a standard deviation 
of $0.025$ in magnitude and color.
The calculations show that the eMSTO and bimodal MS of NGC 1866 cannot be 
reproduced by the effects of the binaries alone. In order to reproduce the eMSTO 
region, an age spread of about $150$ Myr (from 190 to $340$ Myr) 
is required in the simulation. The simulation produced a few BF stars and 
minimal BB stars.

The results cannot exclude the effects of binaries. The mass-ratio $q$ 
is the key parameter determining the evolutions of binaries, so the
calculated results can be affected by the distribution of $q$. 

\subsection{Results for Binary Populations with Different $q$ Distributions}
\subsubsection{The binary population reproducing the BF stars}

\begin{figure}
\flushleft
\includegraphics[scale=0.6, angle=-90]{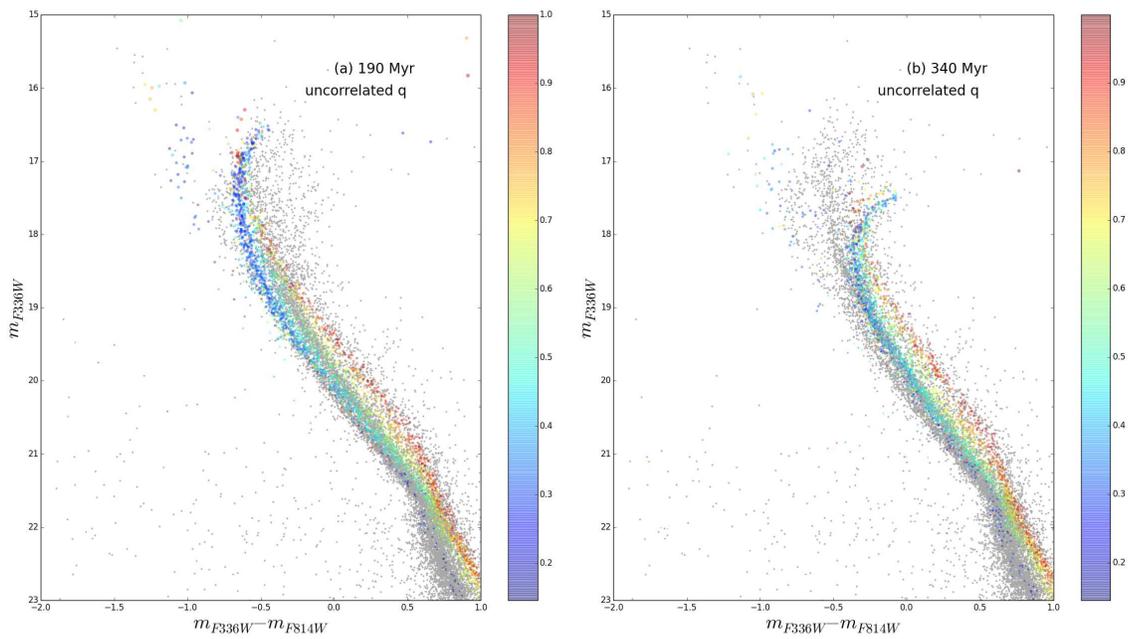}
\caption{Same as Figure \ref{fig2}, but the initial mass of the secondary 
is uncorrelated with that of the primary. In the figure, when the mass of 
the secondary is larger than that of the primary, the value of mass ratio
is redefined as $1/q$. Thus the value of mass ratio keeps within $0-1$ in 
the figure.}  
\label{fig3}
\end{figure}

In order to study whether the BB and BF stars can be reproduced by the evolutions
of binaries, we computed the evolutions of a sample of binaries 
with an uncorrelated $q$. A sample was first
generated at random from the IMF as primary stars. Then secondary stars were
generated at random from the same IMF. We obtained about $3000$ objects
with $15.0\leq m_{F336W}\leq 23.0$ and $-2.0\leq m_{F336W}-m_{F814W}\leq 1.0$
in each simulation, with the merged stars making up around $13\%$.
The CMDs of the simulated population are shown in Figure \ref{fig3}. 
For clarity, when the mass of the secondary star evolved to the given age 
is larger than that of the primary star, the value of the mass ratio 
is redefined as $1/q$ in the Figures. The simulations cannot reproduce enough
BB stars, but they can generate some BF stars (see Figure \ref{fig3}), 
which indicates that the BF stars are relevant to the evolutions of binaries. 

The initial value of $q$ of the simulated BF stars is mainly 
between $3$ and $4$, and the values of initial $M_{1}+M_{2}$ of 
the stars are mainly between about $4$ and $5.5$ \dsm{} for the population
with the age of $340$ Myr. Not all binary systems
with the special $q$ and initial $M_{1}+M_{2}$ between $4$ 
and $5.5$ \dsm{} can evolve into BF stars; only some of the 
binaries can do so. According to the IMF, the more massive the stars, 
the smaller their number. If the secondary mass is determined 
by $qM_{1}$ and the initial $q$ is a uniform 
distribution within $0-1$, the BF stars could hardly be reproduced 
in a cluster because there are not enough binary systems with the initial $q$ 
between $0.25$ and $0.33$ and $M_{1}+M_{2}$ between $4$ and $5.5$ \dsm{}. 
The presence of a large number of BF stars indicates that the initial $q$ 
for the population of NGC 1866 may follow other distributions
or that the BF stars do not belong to NGC 1866.

\begin{figure}
\includegraphics[scale=0.6, angle=-90]{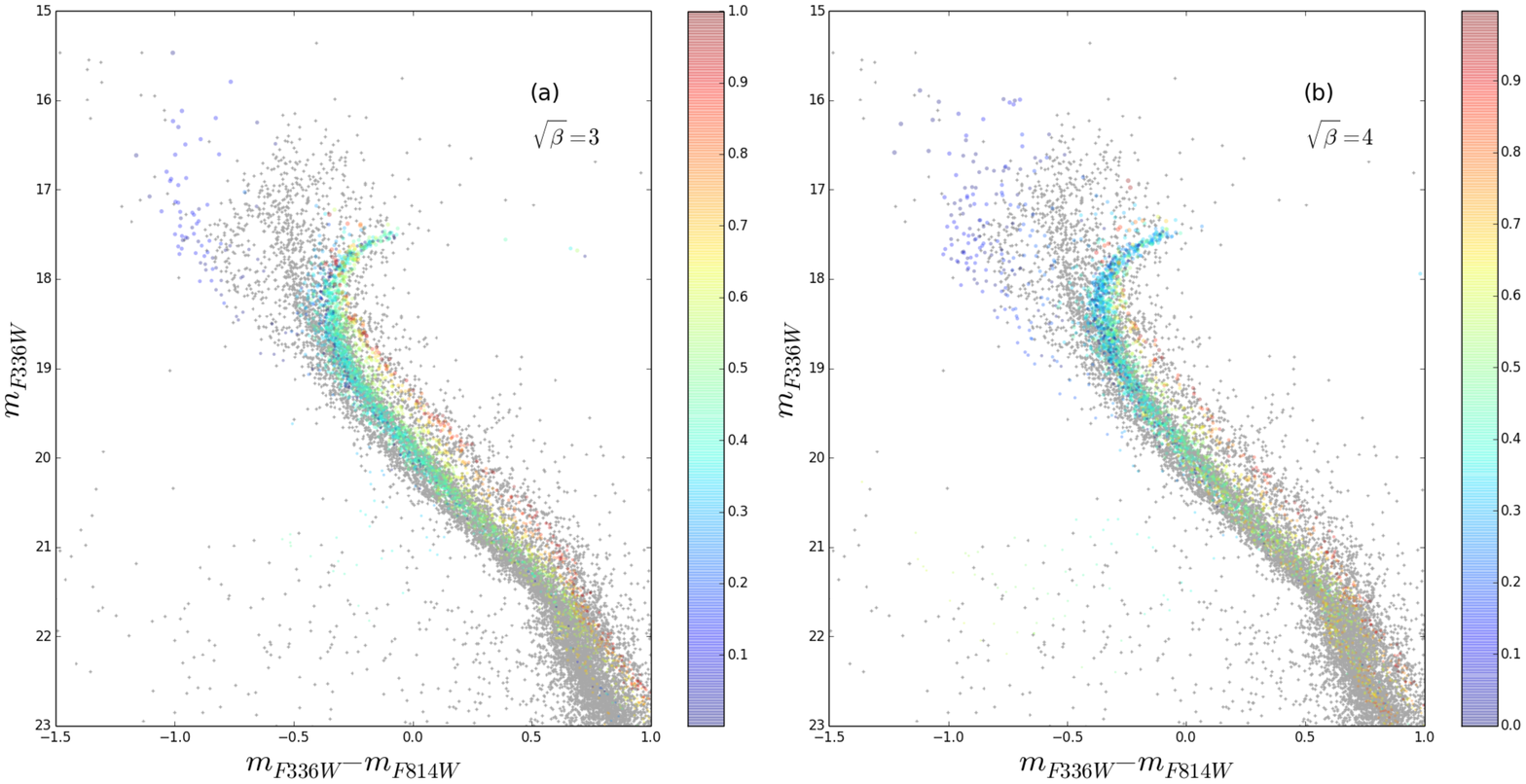}
\includegraphics[scale=0.6, angle=-90]{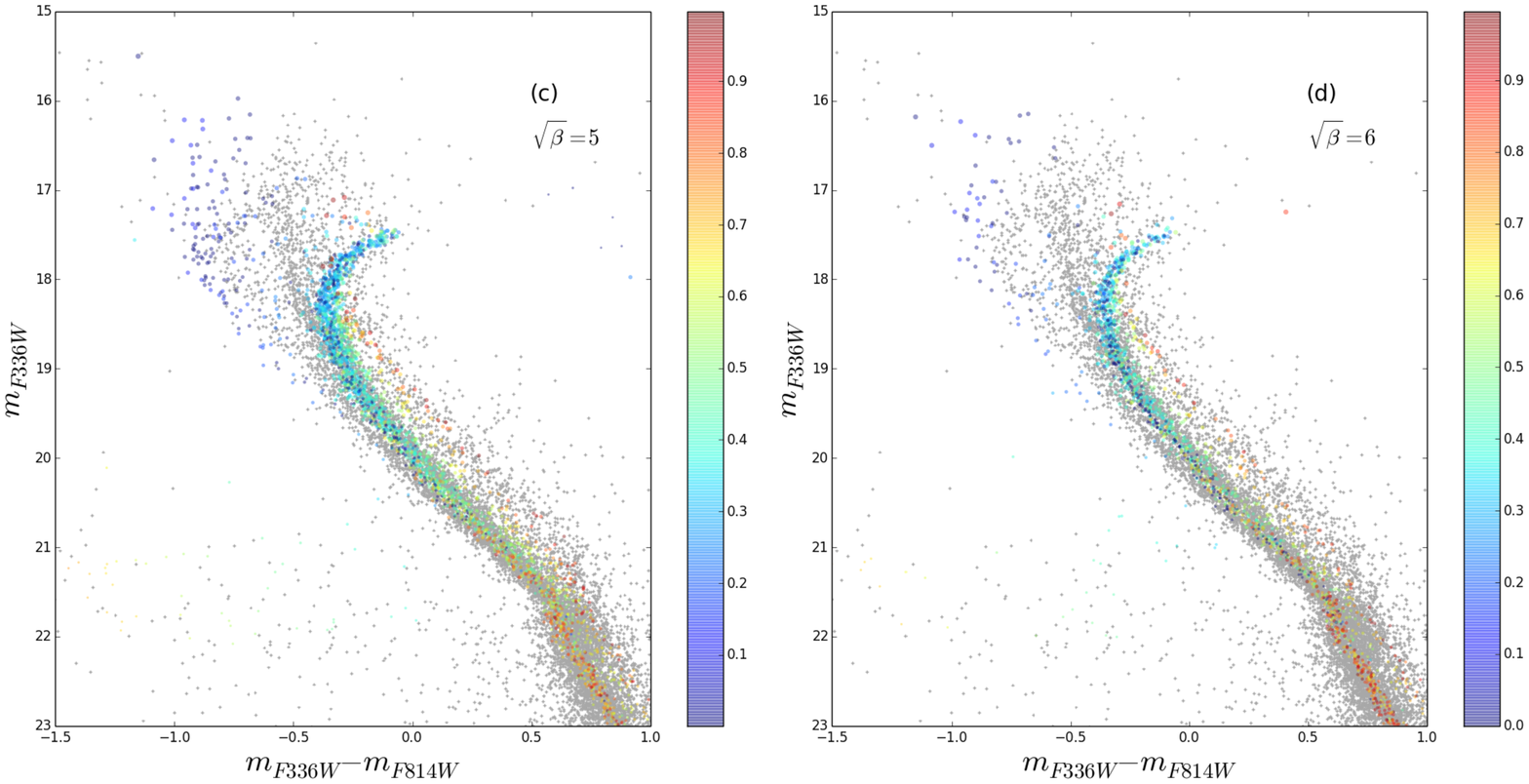}
\caption{CMDs of simulated stars with the age of $340$ Myr but with
different $\beta$. There are about $3000$ simulated objects in panels
$a$ and $b$, $2500$ objects in panel $c$, $1600$ objects in panel $d$. 
The merged stars make up about $13\%$. When the mass of the secondary is 
larger than that of the primary, the value of mass ratio is redefined 
as $1/q$. The larger the $\mathbf{\beta}$, the larger the number of massive
stars. Thus the number of the simulated population decreases. } 
\label{fig4}
\end{figure}

To reproduce the BF stars, the distribution of $q$ is assumed to be
\begin{equation}
 n(q)=\frac{2q}{\mathbf{\beta}},
\end{equation}
where $\mathbf{\beta}$ is a free parameter. Then mass-ratio $q$ 
is generated at random by
\begin{equation}
 q=\sqrt{\mathbf{\beta}}\times\sqrt{r_{i}},
\end{equation}
where $r_{i}$ is a random number within $0-1$. Figure \ref{fig4} represents
the CMDs of simulated populations with different values of $\mathbf{\beta}$
but with the same other parameters, which shows
that the BB and BF stars of NGC 1866 are reproduced well by the simulations
with $\sqrt{\mathbf{\beta}}= 4$ or $5$ (see panels $b$ and $c$ of Figure \ref{fig4}).

When $\sqrt{\beta} \lesssim 4$, the number of simulated BB and BF stars increases 
with the increase in $\sqrt{\beta}$. The smaller the $\sqrt{\beta}$, the redder the 
produced BF stars. The BF stars of NGC 1866 cannot be reproduced by the simulations
with $\sqrt{\beta} < 4$. The larger the value of $\sqrt{\beta}$, the more
massive the star produced by $qM_{1}$. When the value of $qM_{1}$ is larger than 
a certain value, the star evolves to a later stage rather than H-MS or He-MS stage
at the age of $340$ Myr. Thus, when $\sqrt{\beta} \gtrsim 5$, the number of 
simulated BB, BF, and MS stars decreases with an increase in $\sqrt{\beta}$. 
In order to produce the same number of BF stars, the number of initial models of simulation
with $\sqrt{\beta} = 6$ is $1.5$ times as large as that with $\sqrt{\beta} = 5$.

\begin{figure}
\includegraphics[scale=0.6, angle=-90]{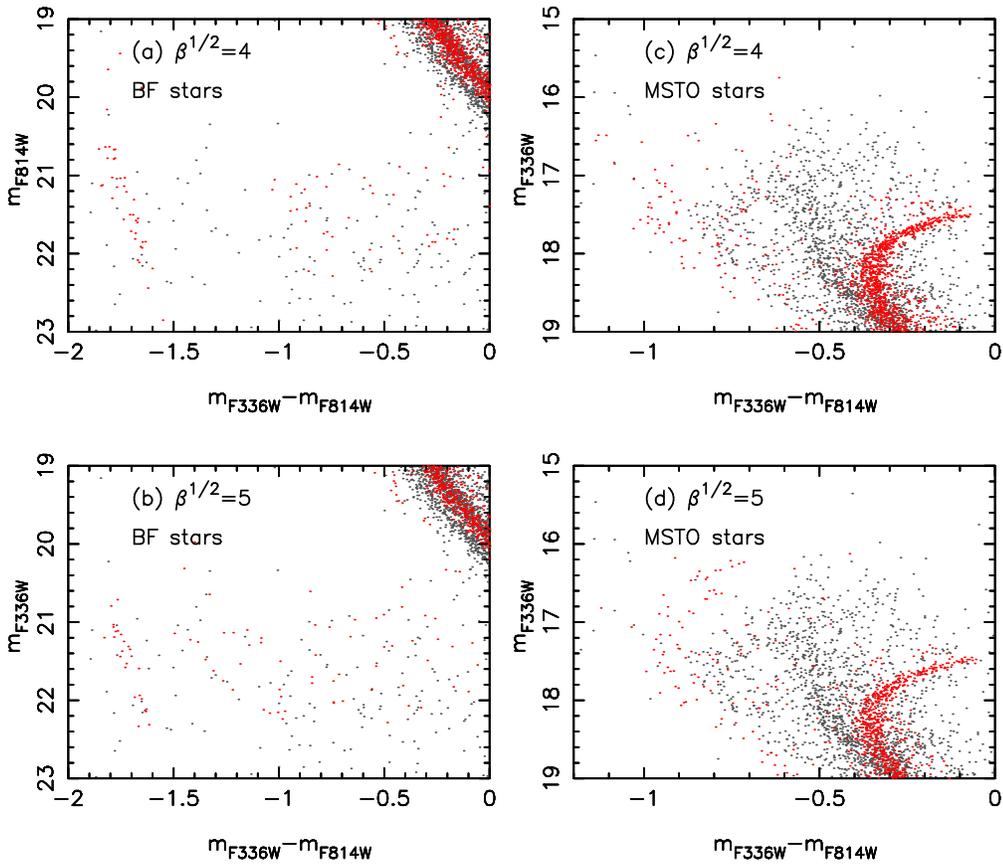}
\caption{CMDs of simulated populations with the age of $340$ Myr but with
different $\mathbf{\beta}$. The grey dots refer to the data of NGC 1866.
The red dots represent the simulated populations.}
\label{fig5}
\end{figure}

The main difference between a simulated population with $\sqrt{\beta} = 4$ and 
that with $\sqrt{\beta} = 5$ is that a simulation with $\sqrt{\beta} = 5$ can 
produce some BF systems with $-1.5 \lesssim m_{F336W}-m_{F814W} \lesssim -1.0$ 
but a simulation with $\sqrt{\beta} = 4$ cannot produce BF systems with 
$-1.5 \lesssim m_{F336W}-m_{F814W} \lesssim -1.0$ (see Figure \ref{fig5}).
Moreover, the number of BB and MS stars of the population with $\sqrt{\beta} = 5$
is slightly lower than that of the population with $\sqrt{\beta} = 4$. 
Furthermore, the simulation with $\sqrt{\beta} = 5$ produces a larger number
of bright stars that deviate from observation than the simulation 
with $\sqrt{\beta} = 4$ (see panels $c$ and $d$ of Figure \ref{fig5}). 
The BF stars of NGC 1866 seem to have a gap between 
$m_{F336W}-m_{F814W} \approx -1.0$ and $ m_{F336W}-m_{F814W} \approx -1.4$
(see panel $a$ of Figure \ref{fig5}), which could be used to limit the value of $\beta$.

Fixing the value of $\sqrt{\mathbf{\beta}}$ at $4$, we computed stellar populations
with different ages. The results are represented in 
Figure \ref{fig6}. The calculations show that the BF stars and the red 
and faint MSTO of NGC 1866 are reproduced well by the populations with 
an age between about $320$ and $340$ Myr. The BF stars and the red 
and faint MSTO belong to the same population. There are almost no BF stars 
with $m_{F336W} < \sim21$ for the stellar populations (see panels $c$ and $d$ 
of Figure \ref{fig6}). A large number of BB stars were reproduced
by the calculations as well.

\subsubsection{The blue and faint stars}

\begin{figure}
\includegraphics[scale=0.6, angle=-90]{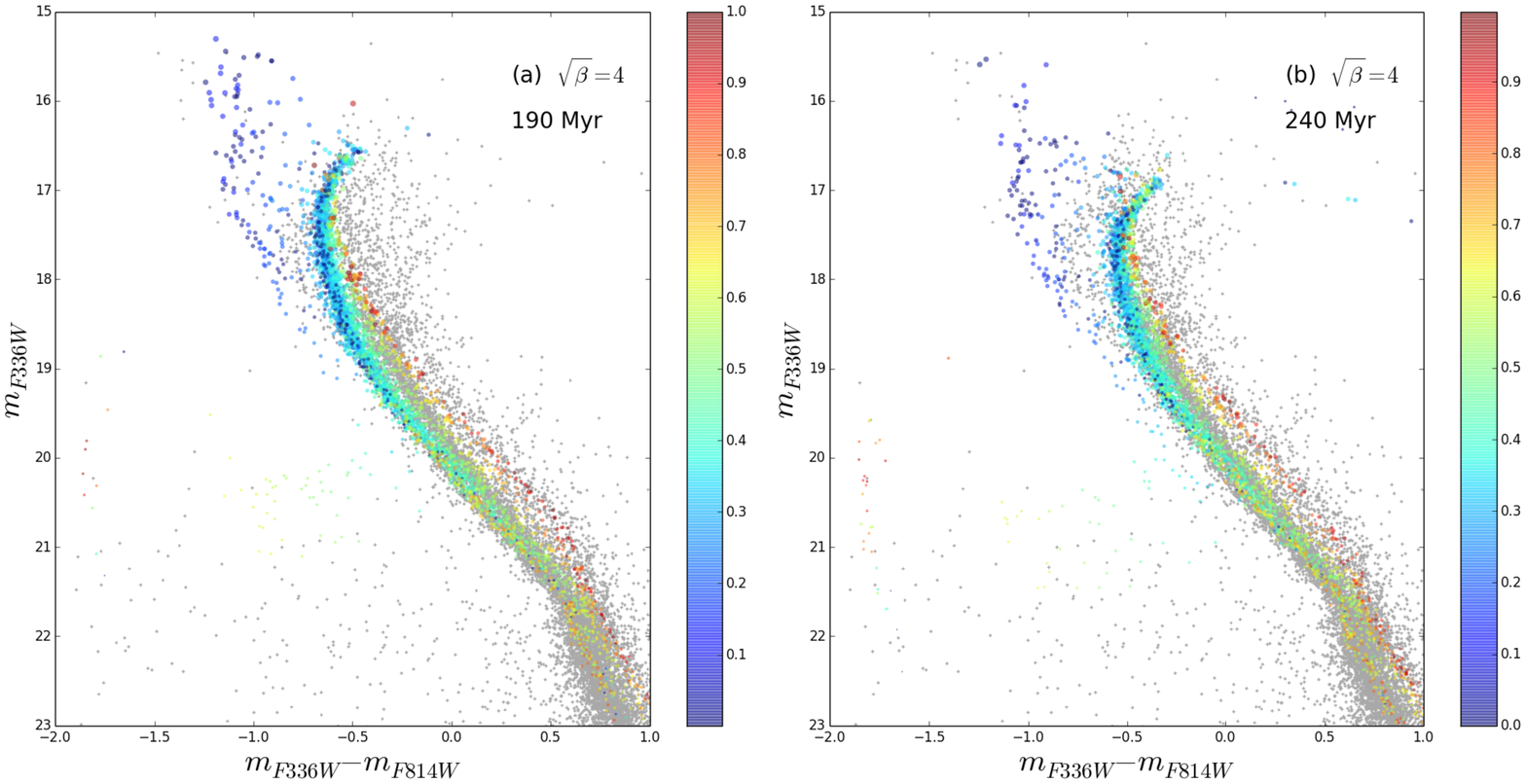}
\includegraphics[scale=0.6, angle=-90]{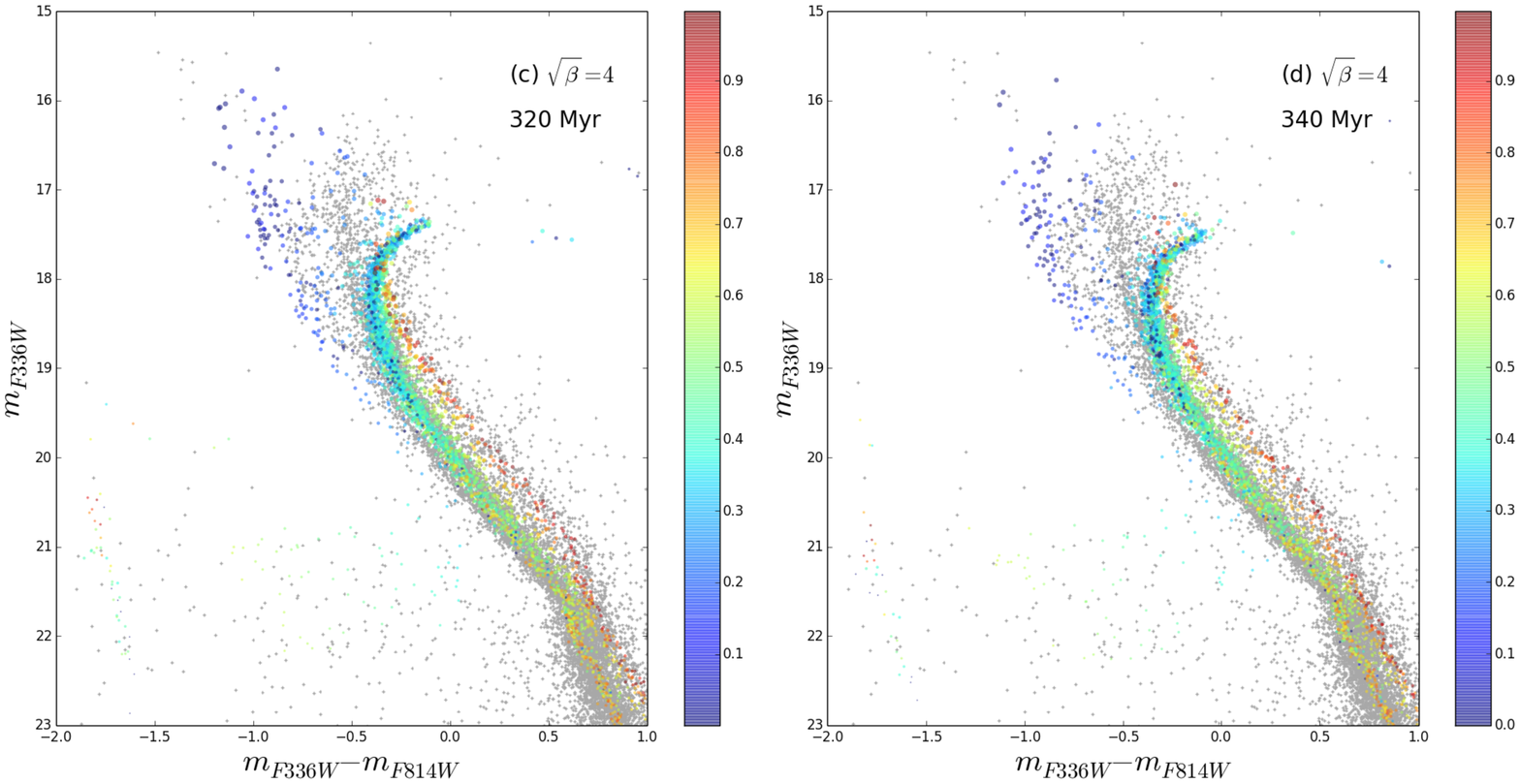}
\caption{Same as Figure \ref{fig4}, but for the populations with different ages.
There are approximately 3000 simulated objects in panels $a$ and $b$ and 
about 2800 simulated objects in panels $c$ and $d$. The simulated stars in the 
lower-left corners are mainly He-star + white dwarf systems.} 
\label{fig6}
\end{figure}

For both the simulated and observed populations, there are only 
a few BF stars brighter than $m_{F336W} \sim 21$ (see
Figure \ref{fig5} and panels $c$ and $d$ of Figure \ref{fig6}), which indicates 
that the luminosity of $m_{F336W}$ for BF stars with the given age has an upper limit.
Our calculations show that a BF star is a binary system consisting of 
a hydrogen MS star and a naked helium star. The initial value of $q$ 
of the system is mainly between about $2.5$ and $4$. The He stars 
evolve from the massive stars of the systems. The initial masses 
of the massive stars are mainly in the range of $\sim 3.0 - 3.5$ \dsm{}
for the population of the age of $340$ Myr. More massive stars 
have evolved into later phases, but lower mass stars have not yet
evolved into He stars. 

\begin{figure}
\centering
\includegraphics[scale=0.6, angle=-90]{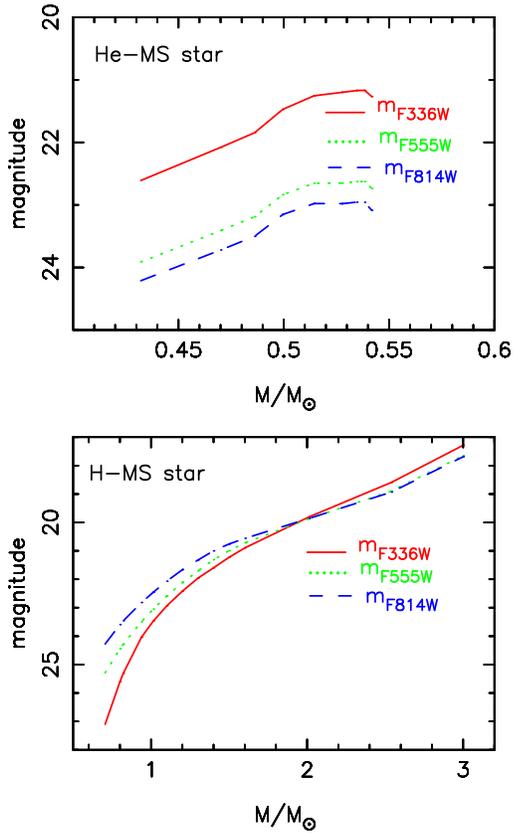}
\caption{ The distributions of magnitudes of stars with the age of $340$ Myr as
a function of mass. }
\label{fig7}
\end{figure}

The masses of the He stars of BF systems are mainly between 
about $0.45$ and $0.54$ \dsm{}, but the masses of the H-MS stars 
are mainly in the range of $\sim 0.7 - 1.3$ \dsm{}. The value 
of $m_{F336W}$ of a BF system is mainly dependent on its He star 
because the value of $m_{F336W}$ of its H-MS star is much 
larger than that of the He star when the mass of the H-MS star 
is lower than $1.4$ \dsm{} (see Figure \ref{fig7}).
The luminosity of a He star is determined by its mass.  
The value of $m_{F336W}$ of the He-MS stars with masses in the range 
of $\sim 0.50 - 0.54$ \dsm{} is around $21$. More massive He stars have 
evolved into the Hertzsprung Gap or later phases. The lifetimes of the
phases are very short, which leads to the fact that there are only a few
BF stars with $m_{F336W} <$ $\sim21$ in our simulations. 
Therefore, the BF candidates cannot blend with BB stars in the CMD
unless the number of H-MS + He-star systems is large enough to contain many 
more massive He stars at the given age.

\begin{figure}
\centering
\includegraphics[scale=0.4, angle=-90]{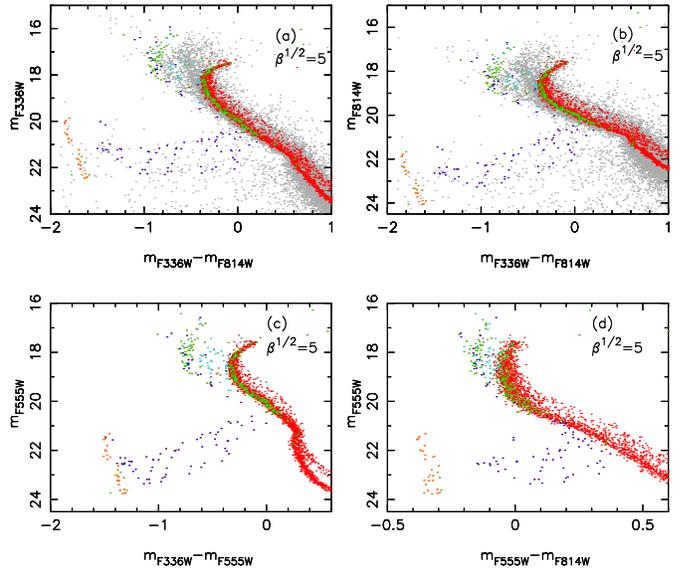}
\caption{CMDs of a simulated population with the age of $340$ Myr. 
There are about $2300$ simulated objects in each panel.
The merged stars (green dots) make up about $13\%$. The four green dots in the 
lower-left corners of the panels are merged He stars. Other green dots are H-MS stars.
The blue and the orange dots represent H-MS + He-star and WD + He-star systems, respectively. 
The cyan dots denote H-MS + WD systems.} 
\label{fig8}
\end{figure}

The luminosities of $m_{F814W}$ of the BF systems are mainly dependent 
on the H-MS stars of the systems because the values of $m_{F814W}$ 
of the H-MS stars are much lower than those of the He stars (see Figure \ref{fig7}).
But the values of $m_{F336W}$ of the systems only slightly decrease 
with an increase in mass of the H-MS stars. Thus the variation in mass 
of the H-MS stars mainly affects the values of $m_{F336W}-m_{F814W}$. 
For the BF stars, the more massive the H-MS stars, the lower the values 
of $m_{F814W}$, and the larger the values of $m_{F336W}-m_{F814W}$;
but the more massive the He stars, the lower the values of $m_{F336W}$. 
The characteristics of the BF stars of NGC 1866 are consistent with 
those of H-MS + He-star systems (see Figure \ref{fig8}).

However, when the H-MS star is massive than about $1.4$ \dsm{},
the luminosity of the H-MS + He-star system can be significantly 
affected by the H-MS star. When the H-MS star is massive than 
about $1.5$ \dsm{}, the values of $m_{F336W}$ and $m_{F814W}$ of the system 
are mainly determined by the H-MS star. The system appears as a MS star 
rather than a BF star. As a consequence, there is an upper limit 
of $m_{F336W}$ for BF stars with the age of $340$ Myr at $m_{F336W} \sim 21$. 
The increase in mass of the H-MS stars of BF systems cannot lead to 
the fact that BF candidates blend with the BB stars in the CMD.
The BF stars with an approximate $m_{F336W}$ are almost in line 
in the CMD due to the difference in mass of H-MS stars (see 
panels $c$ and $d$ of Figure \ref{fig6}). These characteristics 
can be used to restrict the age of young star clusters.

If the age of the BF stars is younger than about $320-340$ Myr, 
the value of the upper limit of $m_{F336W}$ will be smaller than 
$21$ due to the presence of more massive He stars (see panel $a$ of 
Figure \ref{fig6}). In the observed sample, there are only several BF stars 
with $m_{F336W}$ between about $20$ and $21$, but a large number between 
around $21$ and $23$.
Both the BF stars and the red and faint MSTO stars belong to the same
population. This indicates that there is a stellar population as old 
as $\sim320-340$ Myr in NGC 1866. Moreover, BF stars can only derive 
from binary systems with an initial mass ratio in a narrow range. 
Thus there should be many binaries in NGC 1866.  

The orange dots in the lower-left corners of the panels of Figure \ref{fig8}
represent He-star + white dwarf (WD) systems. The magnitudes of these systems are almost entirely 
determined by their He stars. Thus they look like a single He star and have an approximately
equal $m_{F336W}-m_{F814W}$ or $m_{F555W}-m_{F814W}$. Figure \ref{fig7} shows 
that the longer the effective wavelength or the more massive the H-MS star, 
the more easily is the magnitude of H-MS + He-star system affected by 
the H-MS star. Therefore, the H-MS + He-star systems are more easily separated 
from MS by $m_{F336W}-m_{F814W}$ rather than by $m_{F555W}-m_{F814W}$. 
The lower the mass of a H-MS star, the smaller is the contribution of the 
H-MS star to the $m_{F336W}$ and $m_{F555W}$ of the H-MS + He-star system;
so the BF systems with lower-mass H-MS stars are closer to WD + He-star systems 
in $m_{F336W}-m_{F555W}$ (see panel $c$ of Figure \ref{fig8}). This indicates
that BF systems are more easily displayed on $m_{F336W}-m_{FxxxW}$, where the 
$xxx$ represents the effective wavelength of other filters, such as $450$, $555$, 
or $814$. It indicates, too, that BF systems have different behaviors in different colors,
which can aid us in confirming the BF systems.  

The simulation with a larger $\beta$ can produce the H-MS + He-star systems
with a lower-mass H-MS star, i.e. bluer BF stars when $\sqrt{\beta} \leq 5$, which 
leads to the fact that simulations with $\sqrt{\beta} =4$ cannot produce the BF 
stars with $m_{F336W}-m_{F814W} \lesssim -1.0$.

\subsubsection{The blue and bright stars}

The simulations also produce many BB stars that are mainly merged stars, 
H-MS + He-star systems, and H-MS + WD systems. Such stars are also
called blue stragglers \citep{stro70, pols94}.
The H-MS stars had accreted mass from their companions. 
The value of initial $q$ of the systems is mainly in the 
range of around $1.5 - 2.5$. As with BF stars, 
the initial masses of the massive stars that evolved into the naked 
He-MS stars are mainly between about $3.0$ and $3.5$ \dsm{};
but the initial masses of the H-MS stars are higher than those of 
BF systems. The masses of the H-MS stars of BB systems are mainly between 
about $2.5$ and $4$ \dsm{}, which are obviously more massive than 
those of BF systems. The luminosities of the BB systems are more dependent
on their H-MS stars than on their He stars. 
The BB stars thus look like MS stars and are more luminous than BF systems.

The accretion and merging make the stars bluer and brighter than 
the MSTO stars. The value of mass accretion rates can affect the 
luminosity of stars, but the mass accretion rate and the process 
of merging are not definitely known in the theory of stellar evolution.
The more the mass accreted by a star, the larger is the star's luminosity. 
The uncertainty of the rate could give rise to the fact that the simulated 
BB stars are more scattered than the observed ones in the CMD (see 
Figures \ref{fig6} and \ref{fig8}).

The contribution of the He stars to the $m_{F336W}$ of the systems 
is larger than to the $m_{F814W}$ of the systems. The radius of the He 
stars is of the order of $0.1$ \dsr{}. The He stars of the systems could 
be outshone by their companions or disks and be difficult to observe.
This may lead to the fact that the observed BB stars are slightly fainter and redder
than the theoretical models in the CMD.

Some BB stars are H-MS + WD systems. 
The masses of the H-MS stars are mainly between about $3$ and $4$ \dsm{}, 
but those of the WDs are mainly in the range of $\sim 0.55 - 0.7$ \dsm{}.
The luminosities of the systems are mainly determined by the H-MS stars.
Thus they look like blue MS stars rather than WDs.

\subsubsection{The blue MS}
Panel $a$ of Figure \ref{fig6} shows that the simulated stellar population 
with the $q$ distribution and age of $190$ Myr cannot 
reproduce the blue MS of NGC 1866. Moreover, the population contains 
many BF stars with $m_{F336W}$ between about $20$ and $21$. 
The absence of BF stars with $m_{F336W}$ between about $20$ 
and $21$ in the observed sample indicates that properties 
of the blue MS stars of NGC 1866 should be different from those 
of the simulated population.

\begin{figure}
\flushleft
\includegraphics[scale=0.6, angle=-90]{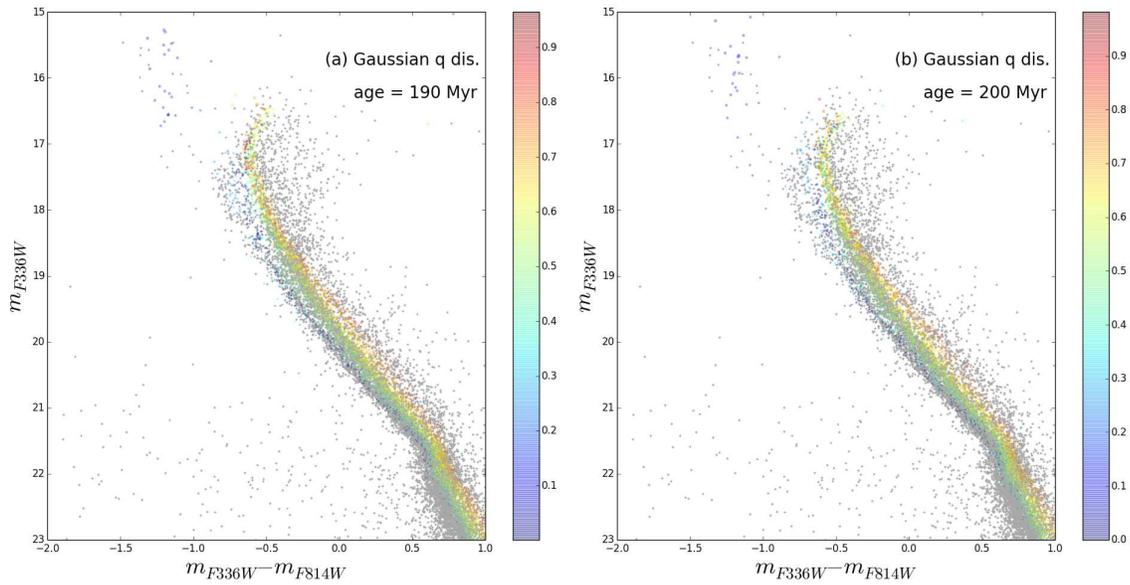}
\caption{Same to Figure \ref{fig2}, but for the population
with the Gaussian $q$ distribution. There are about 
4000 simulated objects in each panel. The merged stars make
up about $12\%$. Others are binaries.} 
\label{fig9}
\end{figure}

\begin{figure}
\centering
\includegraphics[scale=0.6, angle=-90]{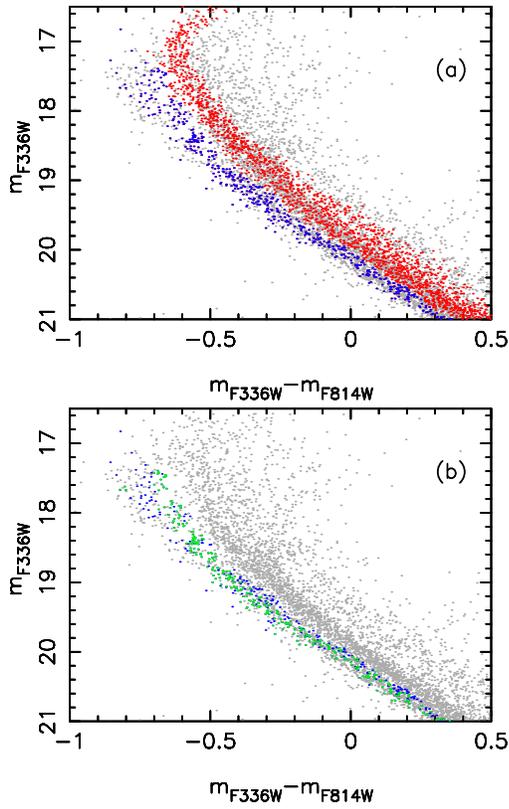}
\caption{Enlarged panel $a$ of Figure \ref{fig9}. The simulated
blue and red MS stars are separated by a visual check. The blue dots
represent the blue MS of the simulated population in panel $a$. In panel $b$,
the green dots show the merged (single) stars and make up $57\%$ of the simulated
blue MS, while blue dots represent binaries. } 
\label{fig10}
\end{figure}

We computed the evolutions of binary populations with a Gaussian mass-ratio
distribution. The mean value and standard deviation of the Gaussian distribution
is $0.6$ and $0.11$, respectively. Figures \ref{fig9} and \ref{fig10} show that 
the bimodal MS and blue MS of NGC 1866 can be reproduced by 
the binary population with an age of about $190$ Myr. The blue stragglers 
with $m_{F336W} \sim 16$ and $m_{F336W}-m_{F814W}< -1.0$ of NGC 1866 
are reproduced as well. As we have just said, however, the BF stars 
do not appear in this simulation; so this young population is not incompatible 
with the constraints of the observed BF stars.

\begin{figure}
\centering
\includegraphics[scale=0.7, angle=-90]{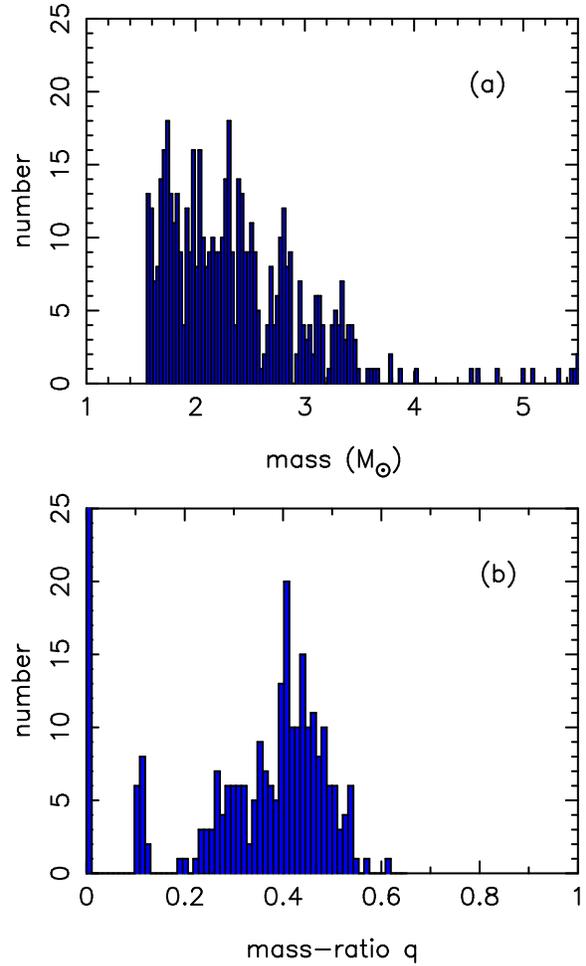}
\caption{Histograms of mass and mass-ratio $q$ of the simulated blue MS stars shown in
panel $a$ of Figure \ref{fig10}. }
\label{fig11}
\end{figure}

The blue MS is mainly composed of merged stars and binaries with a 
$q$ less than about $0.5$ (see Figure \ref{fig11}). 
Most of the simulated systems have an initial 
$q$ larger than $0.5$, which leads to the fact that the red MS is denser 
than the blue MS. Mass accretion and merging can lead to the fact that 
the number of stars with masses in a certain range can increase, 
but those with masses in another range decreases (
see Figure \ref{fig11}).
This results in the fact that the blue MS is discontinuous in the CMD. 

The simulated blue stragglers are composed of merged MS stars and H-MS 
+ He-star systems. The masses of the He-MS stars are mainly between 
about $0.6$ and $0.7$ \dsm{}. As we have noted, the H-MS stars
had accreted mass from their companions. The masses of these stars 
and the merged MS stars are mainly in the range of $\sim 5.0$ and $6.5$ \dsm{}, 
which is much larger than the masses of the MSTO stars of the SSP with the age of $190$ Myr.
The luminosity of the H-MS + He-star system is mainly dependent on the H-MS star 
rather than the He star. Thus the system is a blue and bright star 
rather than a blue and faint star.

Moreover, part of the simulated blue MS ($m_{F336W} < 19$) 
are merged MS stars, H-MS + He-star systems, and H-MS + WD systems.
The masses of the He-MS stars are mainly between about $0.6$ and $0.7$ \dsm{};
but those of the WDs mainly between the range of $\sim 0.9 - 1.0$ \dsm{}. 
For this young stellar population, the masses of the He stars of H-MS + He-star systems
are mainly between about $0.6$ and $0.7$ \dsm{}, which are larger than 
those of the stellar population with the age of $340$ Myr.

\begin{figure}
\includegraphics[width=7cm, angle=-90]{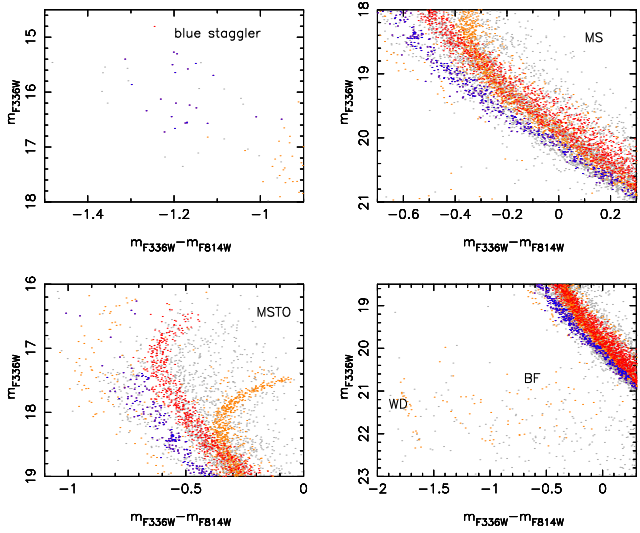}
\caption{CMDs of multiple stellar populations. The red and blue dots
represent the binary population with the Gaussian mass-ratio distribution
and with the age of $190$ Myr, while the orange dots shows the population with 
$\sqrt{\beta}=5$ and with the age of $340$ Myr. }
\label{fig12}
\end{figure}

Figure \ref{fig12} presents the CMD of simulated multiple populations that are the 
mixture of the binary population characterized by Gaussian mass-ratio distribution
and an age of $190$ Myr with the population with $\sqrt{\beta}=5$ 
and an age of $340$ Myr. It shows that the blue and the red MS  
gradually merge when the value of $m_{F336W}$ is larger than $20$. 
When the value of $m_{F336W}$ is less than $20$, however, the separation
between the blue and the red MS is almost not affected by the old population.

\subsubsection{Other Clusters}

\begin{figure}
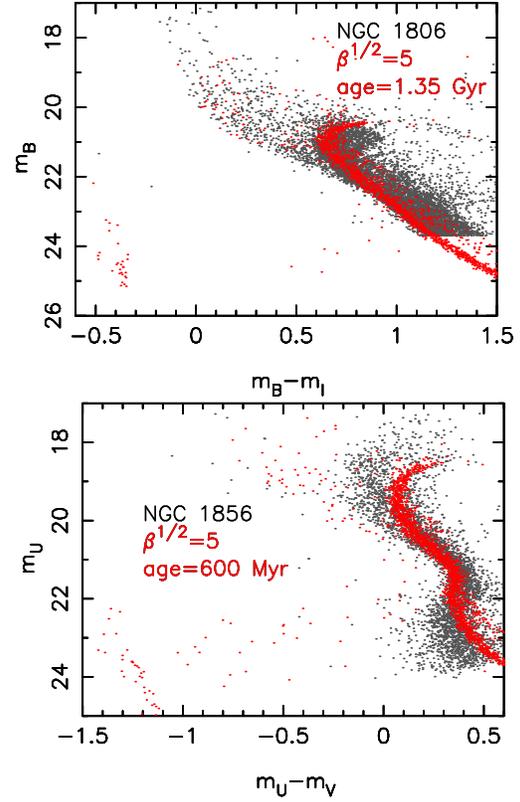

\centering
\includegraphics[scale=0.6, angle=-90]{fig13-1.ps}
\includegraphics[scale=0.6, angle=-90]{fig13-2.ps}
\caption{CMDs of the observed and simulated stellar populations.
The dark gray dots and the red ones represent the observed and the simulated stars,
respectively. There are about 2000 simulated objects in each panel. The values of
distance modulus and reddening $E(B-V)$ are 18.8 and 0.16 for NGC 1806 and 18.5 and 
0.08 for NGC 1856, respectively. } 
\label{fig13}
\end{figure}

The CMDs of NGC 1806 and NGC 1856 have been given by \citep{milo09, milo15}.
There are almost no BF stars in the observed CMDs. With the same initial parameters 
and $\sqrt{\beta}=5$, we computed stellar populations with different ages.
The blue stragglers and the blue and bright MSTO of NGC 1806 can be reproduced by 
a population with the age of $1.35$ Gyr (see Figure \ref{fig13}); but there are only 
a few BF stars in the simulated population. The faint and red MSTO of NGC 1856 can be
reproduced by a population with the age of 600 Myr, and the few BF 
stars of NGC 1856 also are reproduced by the simulated population. The observations
and simulations show that BF stars are correlated with the age of clusters.
They appear more easily in a young cluster.

\section{DISCUSSION AND SUMMARY}
\subsection{Discussion}

\begin{figure}
\centering
\includegraphics[scale=0.6, angle=-90]{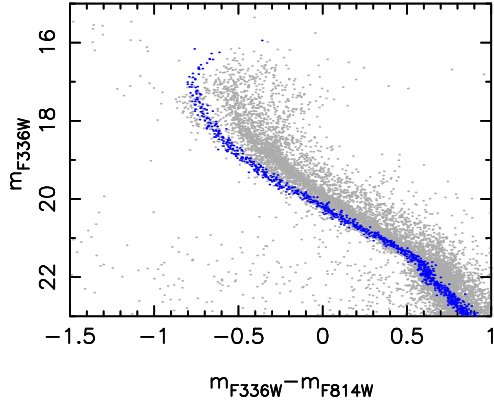}
\caption{The CMD of the simulated SSP (blue dots) with the age of $150$ Myr. }
\label{fig14}
\end{figure}

We noticed that the ages of binary populations are larger than those given 
by \cite{milo17}, which may be due to the effects of binaries. Similar to 
\cite{milo17}, we obtained an age of about $150$ Myr for NGC 1866 when
we used a SSP to fit the blue MS of NGC 1866 (see Figure \ref{fig14}).
Moreover, the simulated age of a cluster is related to [Fe/H]. 
The uncertainty of [Fe/H] can affect the age.

\begin{figure}
\centering
\includegraphics[scale=0.6, angle=-90]{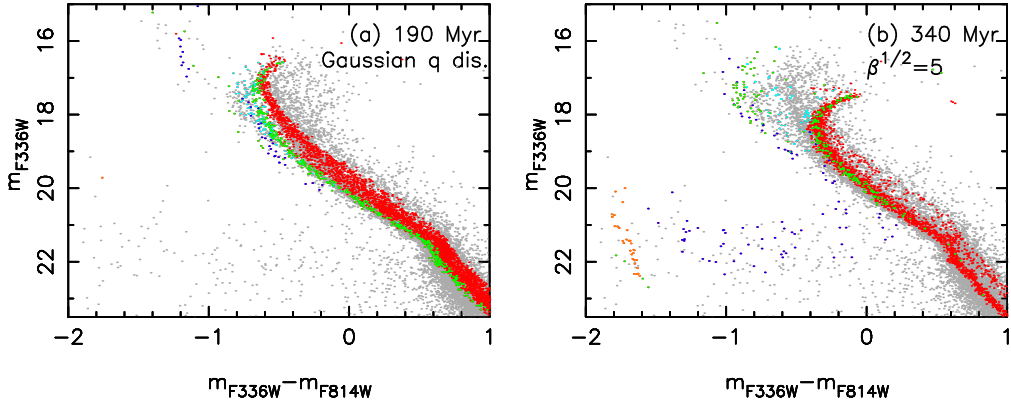}
\caption{ CMDs of stellar populations obtained from \cite{salp55} IMF 
rather than \cite{chab01} IMF. The merged stars (green dots) make up around $14\%$; 
others are binaries. The blue and the orange dots represent H-MS + He-star 
and WD + He-star systems, respectively. The cyan dots denote H-MS + WD systems.} 
\label{fig15}
\end{figure}

Assuming a constant mass-ratio distribution, \cite{milo17} found
that the fraction of binaries in NGC 1866 is about $0.28$.
In our simulations, in the primary stage of evolutions all models are 
members of binaries. When the models are evolved to the age of $190$ 
or $340$ Myr, the merged stars make up only about $12\%-14\%$ of the obtained
objects; but all the others are still binaries. Moreover, all the simulated BF stars are binaries.
The ratio of merged He stars to He-star + WD systems in Figures \ref{fig8}
and \ref{fig15} is around $0.08-0.12$. Our binary fraction is much larger than 
that given by \cite{milo17}.

The IMF does not affect our results. Basing on \cite{salp55} IMF,
we obtained almost the same results (see Figure \ref{fig15}). The difference
between the results obtained from \cite{salp55} IMF and those based on \cite{chab01} IMF
can be neglected. The value of $E(B-V)$ is $-0.01$ for NGC 1866, $0.08$ for NGC 1856, 
and $0.16$ for NGC 1806, which may be related to the mixing-length 
parameter. That parameter is unadjustable in the \cite{hurl02}
codes and is calibrated to a solar model. The larger the parameter,
the smaller the radius of stellar models, and the higher their effective 
temperature. If the value of the mixing-length parameter increases with an 
increase in the mass of stars but is fixed at the value calibrated 
to a low-mass star, one could find that for clusters with the same reddening  
the value of theoretical $E(B-V)$ decreases with a decrease
in age of the clusters because the masses of MSTO stars increase
with the decrease in age of the clusters.

\begin{figure}
\flushleft
\includegraphics[scale=0.6, angle=-90]{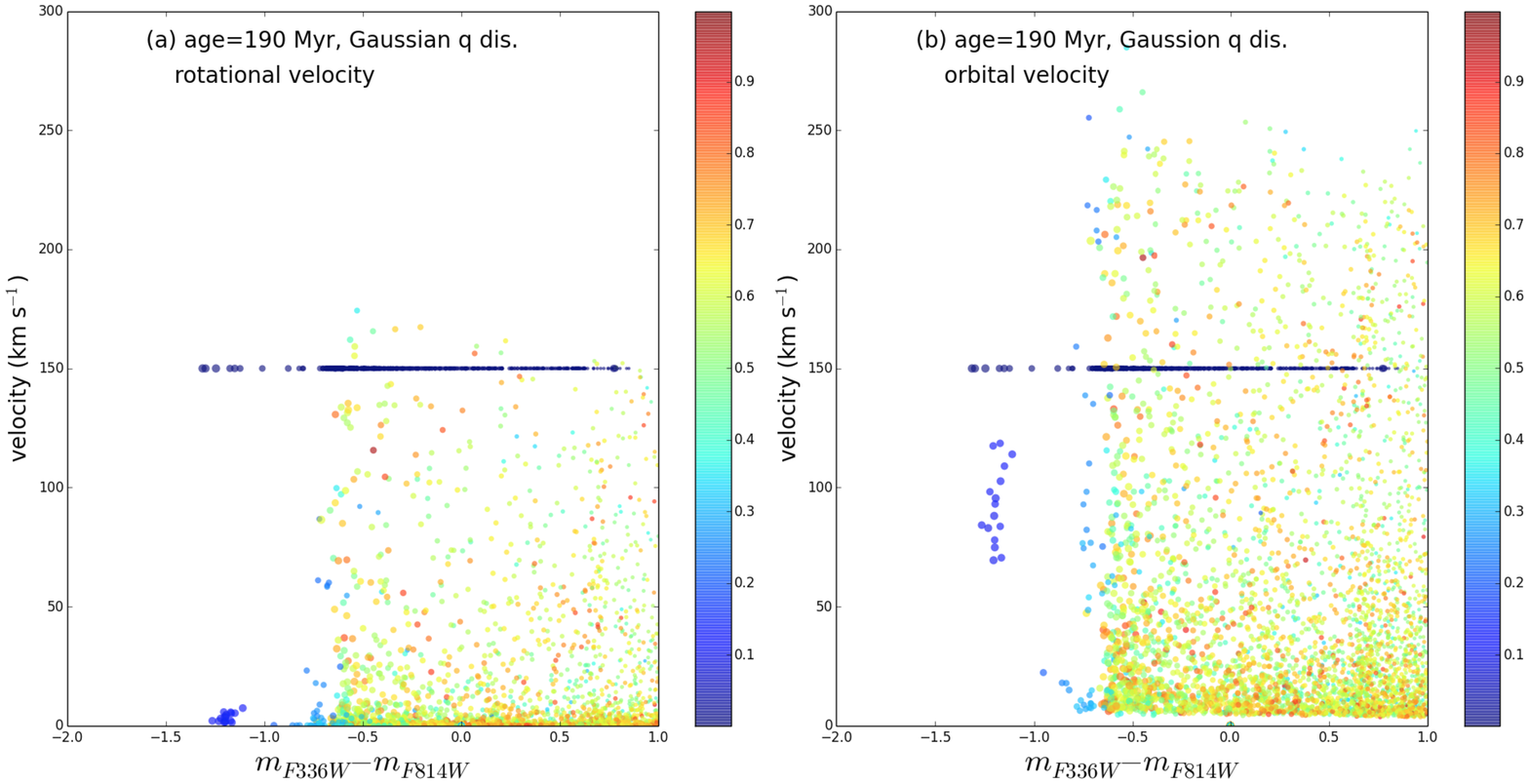}
\includegraphics[scale=0.6, angle=-90]{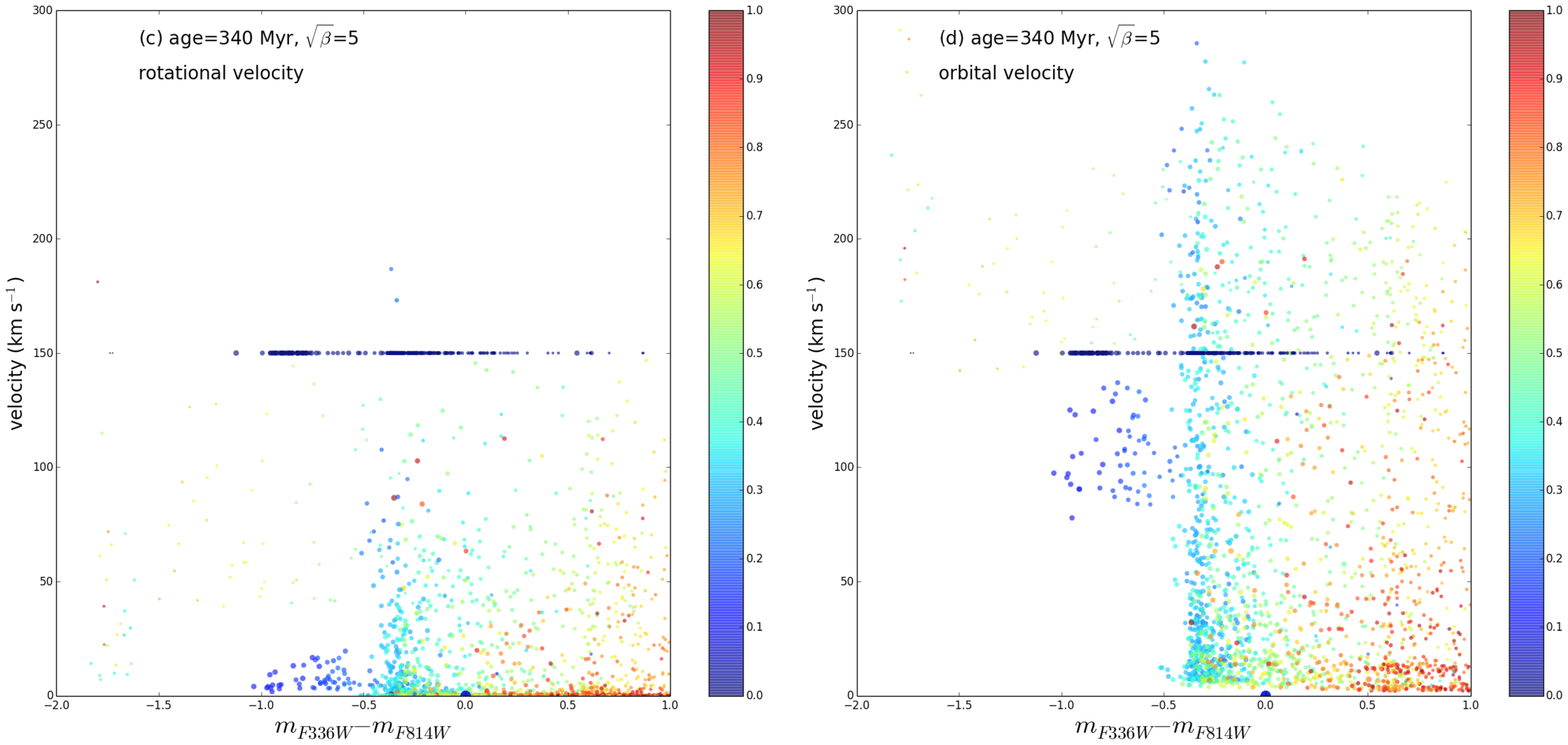}
\caption{Rotational velocities and orbital velocities of simulated populations 
as a function of $m_{F336W}-m_{F814W}$. The size and the color of the dots are 
proportional to stellar mass and mass-ratio $q$, respectively. The velocities
of merged stars are assumed to be $150$ km s$^{-1}$.} 
\label{fig16}
\end{figure}

The rotation velocity of merged stars is hard to estimate. If they 
lost angular momentum as they merged, they are slowly rotating stars;
but if their angular momentum was conservative, they are fast rotators.  
In order to distinguish the merged stars from binaries in Figure \ref{fig16}, 
their velocities are assumed to be $150$ km s$^{-1}$. We calculated
rotational velocities and orbital velocities with the assumption 
that the rotation rate of a star is equal to the revolution rate of 
the binary system. The rotational and orbital velocities of many stars 
are of the order of $100$ and $200$ km s$^{-1}$, respectively; but 
orbital velocities can be as high as about $300$ km s$^{-1}$.
The fraction of MSTO stars with orbital velocities higher than $200$ km
s$^{-1}$ in the old population is larger than that in the young population
(see panels $b$ and $d$ of Figure \ref{fig16}).

The eMSTO and bimodal MS of NGC 1866 can be explained by a combination
of rotation and age spread \citep{milo17, corr17}; but the eMSTO can also
be explained by the effects of a large overshoot of the convective core of stars
\citep{yang17}. A single He star, WD, or MS star cannot appear
as a BF star, BB star, or blue straggler. Thus the H-MS + He-star systems and H-MS + WD 
systems are the main characteristics that can distinguish the binary scenario
from the rotation scenario and the overshoot scenario. 

The calculations show that the BF stars are H-MS + He-star systems.
The H-MS star is a low mass star, and it hardly accretes mass from its
companion. The BF stars and the red and faint MSTO belong to the 
same population. The BF stars can provide a constraint on the age 
of the stellar population. Once the BF stars are determined 
to be the members of the cluster, the existence of an old stellar
population in the cluster will be confirmed. The initial $q$ 
of the H-MS + He-star systems is in a narrow range. The observed 
BF stars could be used to estimate the fraction of binaries in the cluster. 

The simulated bimodal MS is sensitively dependent on the mean value of the 
Gaussian mass-ratio ($q$) distribution and can be reproduced when the mean value 
is in the range of about $0.6-0.67$. 

For a population with an age of $340$ Myr, the masses of He stars 
and WDs are mainly within the range of $\sim0.46-0.54$ and 
$\sim0.55-0.7$ \dsm{}, respectively; but for that with 
an age of $190$ Myr, their masses are mainly within
the range of $\sim0.6-0.7$ and $\sim0.9-1.0$ \dsm{}, respectively.
The fundamental parameters of binary stars are more easily determined from 
light curves than those of single stars. If the young cluster consists 
of two main populations, there would be a bimodal distribution of the
mass of He stars.

Furthermore, part of the blue stragglers are massive H-MS + He-star systems.
If they derive from a young population with an age of about $190$ Myr,
the masses of the He stars are between about $0.6$ and $0.7$ \dsm{},
but those of the H-MS stars are between around $5$ and $6.5$ \dsm{}. 
The masses of the H-MS and He stars of the BF systems of the stellar sub-population with 
the age of $340$ Myr are virtually all lower than $1.4$ and $0.54$ \dsm{}, respectively. 
If one can determine that part of the blue stragglers are the massive
H-MS + He-star systems and that the BF stars are the low mass H-MS + He-star 
systems, that will confirm the existence of the eSFH and the role of binaries. 

\begin{figure}
\flushleft
\includegraphics[scale=0.6, angle=-90]{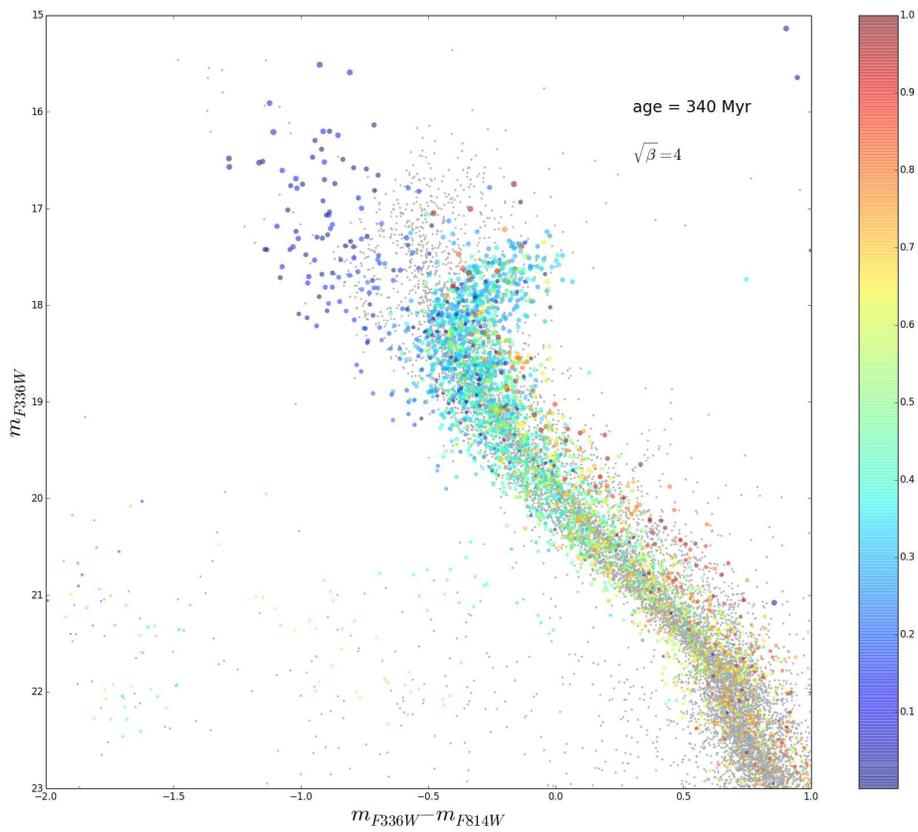}
\caption{\textbf{Same as Figure \ref{fig4}, but photometric errors for artificial
stars are taken to be a Gaussian distribution with a mean value of $0$ and 
a standard deviation of $0.1$ mag in magnitude and color.}} 
\label{fig17}
\end{figure}

\textbf{Photometric errors for bright MS stars are around $0.02$ mag in color 
\citep{corr14, milo16}. The errors could increase with an increase in magnitude.
The characteristics of the BF population are mainly dependent on He stars. 
The values of $m_{F336W}-m_{F814W}$ of the BF population are mainly between 
about $-2.0$ and $-0.3$, i.e. the change of $|m_{F336W}-m_{F814W}|$ is 
large than $0.1$. Even photometric errors increase from $0.02$ to $0.1$ mag,
our results for the BF population are not changed (see Figure \ref{fig17}); 
but the simulated MS broadens.
}

\subsection{Summary}
In this work, we calculated different binary populations. The calculations
show that the BF stars of NGC 1866 are H-MS + He-star systems, which 
derive from the evolutions of binaries with the initial $q$ in the 
range of about $2.5-4$. The He star comes into being from the 
evolution of the massive star of the system. The value of $m_{F336W}$ 
of a BF star is mainly dependent on the mass of the He star. 
The mass of most of the He stars is lower than $0.54$ \dsm{} 
for the population with the age of $340$ Myr, which leads to the 
fact that the $m_{F336W}$ of most of the BF stars is larger than
$\sim 21$. The value of $m_{F336W}-m_{F814W}$ of a BF star 
mainly depends on the mass of the H-MS star that lies mainly within 
the range of about $0.7-1.3$ \dsm{}. The more massive the H-MS star, 
the larger the value of $m_{F336W}-m_{F814W}$ for a BF star. However,
when the mass of the H-MS star is larger than $1.5$ \dsm{}, the 
system looks like an MS star rather than a BF star.
If the population of BF stars can be confirmed to be members of NGC 1866, 
this would directly show that NGC 1866 hosts a population
older than the blue MS, otherwise a mechanism making some stars bluer 
and brighter, such as variable overshoot, is required to explain the blue MS. 

The bimodal MS of NGC 1866 can be produced by the binary population with
the Gaussian $q$ distribution and the age of $190$ Myr. The calculations
show that the BB stars of NGC 1866 are mainly merged stars, H-MS + He-star 
systems, and H-MS + WD systems, which leads to the discontinuities between
the BB stars and the blue MS. The H-MS + He-star systems and H-MS + WD systems
are the main characteristic of the binary models, and this can be used to
confirm or exclude the binary scenario.

Moreover, the calculations show that the blue stragglers of NGC 1866 and 
the blue MS belong to the same population. The blue stragglers consist of 
the merged stars and H-MS + He-star systems. The masses of the He stars are 
mainly between about $0.6$ and $0.7$ \dsm{}, which are larger than 
those of the BF stars. Therefore, the existence of the blue stragglers
and the BF stars can confirm the existence of the eSFH.

To explain the eMSTO and bimodal MS of NGC 1866, a combination of an age 
spread and binary population is required. The role of binaries in the formation
of the eMSTO and bimodal MS can be tested by whether part of the BB stars are
H-MS + He-star systems or H-MS + WD systems. Moreover, the eSFH can be
confirmed by whether the BF stars are members of NGC 1866, because 
the BF stars belong to an older population in the theoretical model.

\acknowledgments
The author thanks the anonymous referee for helpful comments that
helped the author improve this work, A. P. Milone for providing 
the observed data of NGC 1866, and Daniel Kister for help in improving 
the English, and acknowledges the support from the NSFC 11773005, U1631236, and 11273012.

\clearpage

\clearpage

\clearpage

\clearpage

\clearpage

\clearpage

\clearpage

\clearpage

\clearpage

% \clearpage

\clearpage

\clearpage

\clearpage


\begin{thebibliography}{}
\bibitem[Bastian et al.(2017)]{bast17} Bastian, N., Cabrera-Ziri, I., Niederhofer, F., et al. 2017, \mnras, 465, 4795
\bibitem[Bastian \& de Mink(2009)]{bast09} Bastian, N., \& de Mink, S. E. 2009, MNRAS, 398, L11
\bibitem[Bastian et al.(2016)]{bast16} Bastian, N., Niederhofer, F., Kozhurina-Platais, V., et al. 2016, \mnras, 460, L20
\bibitem[Brandt \& Huang(2015)]{bran15} Brandt, T. D., Huang, C. X. 2015, \apj, 807, 25
\bibitem[Chabrier(2001)]{chab01} Chabrier, G. 2001, ApJ, 554, 1274
\bibitem[Correnti et al(2014)]{corr14} Correnti, M., Goudfrooij, P., Kalirai, J. S.,
et al. 2014, \apj, 793, 121
\bibitem[Correnti et al(2015)]{corr15} Correnti, M., Goudfrooij, P., Puzia, T. H., de Mink, S. E. 2015, \mnras, 450, 3054
\bibitem[Correnti et al(2017)]{corr17} Correnti, M., Goudfrooij, P., Bellini, A., 
Kalirai, J. S. Puzia, T. H. 2017, \mnras, 467, 3628

\bibitem[D'Antona et al.(2015)]{dant15} D'Antona, F., Di Criscienzo, M., Decressin, T., et al. 2015, \mnras, 453, 2637
\bibitem[de Grijs(2017)]{degr17} de Grijs, R. 2017, NatAs., 1, 0011
\bibitem[Dupree et al(2017)]{dupr17} Dupree, A. K., Dotter, A., Johnson, C. I., et al. 2017, ApJL, 846, L1
\bibitem[Girardi et al.(2009)]{gira09}
Girardi, L., Rubele, S., \& Kerber, L. 2009, MNRAS, 394, L74
\bibitem[Glatt et al.(2008)]{glat08}
Glatt, K., Grebel, E. K., \& Sabbi, E. et al. 2008, AJ, 136, 1703
\bibitem[Goudfrooij et al.(2014)]{goud14} Goudfrooij, P., Girardi, L., Kozhurina-Platais, V.,
et al. 2014, \apj, 797, 35
\bibitem[Goudfrooij et al.(2011)]{goud11}
Goudfrooij, P., Puzia, T. H., Chandar, R., \& Kozhurina-Platais, V.
2011, \apj, 737, 4
\bibitem[Goudfrooij et al.(2009)]{goud09} Goudfrooij, P., Puzia, T. H., Kozhurina-Platais, V., 
\& Chandar, R. 2009, AJ, 137, 4988
\bibitem[Goudfrooij et al.(2017)]{goud17} Goudfrooij, P., Girardi, L., Correnti, M. 2017, ApJ, 846, 22
\bibitem[Han et al.(1995)]{han95} Han, Z., Podsiadlowski, P., \& Eggleton, P. P. 1995, MNRAS, 272, 800
\bibitem[Hurley et al. (2000)]{hurl00} Hurley, J. R., Pols, O. R., \& Tout, C. A. 2000, MNRAS, 315, 543
\bibitem[Hurley et al. (2002)]{hurl02}
Hurley, J. R., Tout, C. A., \& Pols, O. R. 2002, MNRAS, 329, 897

\bibitem[Lejeune et al.(1998)]{leje98} Lejeune, T., Cuisinier, F., \& Buser, R. 1998, \aap, 130, 65

\bibitem[Lemasle et al.(2017)]{lema17} Lemasle, B., Groenewegen, M. A. T., Grebel, E. K., et al. 2017, A\&A, 608, A85
\bibitem[Li et al.(2014)]{lic14} Li, C., de Grijs, R., Deng, L. 2014, Nature, 516, 367
\bibitem[Li et al.(2017)]{lic17} Li, C., de Grijs, R., Deng, L., \& Milone, A. P. 2017, ApJ, 844, 119
\bibitem[Li et al.(2015)]{liz15} Li, Z., Mao, C., Chen, L. 2015, \apj, 802, 44
\bibitem[Li et al.(2012)]{liz12} Li, Z., Mao, C., Chen, L., Zhang, Q. 2012, \apjl, 761, 22

\bibitem[Mackey \& Broby Nielsen(2007)]{mack07}
Mackey, A. D., \& Broby Nielsen, P. 2007, MNRAS, 379, 151
\bibitem[Mackey et al.(2008)]{mack08}
Mackey, A. D., Broby Nielsen, P., Ferguson, A. M. N., \& Richardson,
J. C. 2008, ApJ, 681, L17
\bibitem[Milone et al.(2009)]{milo09}
Milone, A. P., Bedin, L. R., Piotto, G., \& Anderson, J. 2009, A\&A, 497, 755
\bibitem[Milone et al.(2013)]{milo13} Milone, A. P., Bedin, L. R., Cassisi, S., Piotto, G.,
Anderson, J., Pietrinferni, A., Buonanno, R. 2013, A\&A, 555, A143
\bibitem[Milone et al.(2015)]{milo15} Milone, A. P., Bedin, L. R., Piotto, G., et al. 2015, \mnras, 450, 3750
\bibitem[Milone et al.(2016)]{milo16} Milone, A. P., Marino, A. F., D'Antona, F., et al. 2016, \mnras, 458, 4368
\bibitem[Milone et al.(2017)]{milo17} Milone, A. P., Marino, A. F., D'Antona, F., et al. 2017, MNRAS, 465, 4363

\bibitem[Niederhofer et al.(2015a)]{nied15a} Niederhofer, F., Georgy, C., Bastian, N., Ekström, S.
2015a, \mnras, 453, 2070
\bibitem[Piatti \& Cole(2017)]{piat17} Piatti, A. E., \& Cole, A. 2017, MNRAS, 470, L77
\bibitem[Pols \& Marinus(1994)]{pols94} Pols, O. R., \& Marinus, M. 1994, A\&A, 288, 475
\bibitem[Rubele et al.(2010)]{rube10}
Rubele, S., Kerber, L., \& Girardi, L. 2010, MNRAS, 403, 1156
 
\bibitem[Salinas et al.(2016)]{sali16} Salinas, R., Pajkos, M. A., Strader, J., Vivas, A. K.,
Contreras Ramos, R. 2016, ApJL, 832, L14
\bibitem[Salpeter(1955)]{salp55} Salpeter, E. E. 1955, ApJ, 121, 161
\bibitem[Strom \& Strom(1970)]{stro70} Strom, K. M., \& Strom, S. E. 1970, ApJ, 162, 523
\bibitem[Yang (2016)]{yang16} Yang, W. 2016, \apj, 821, 108
\bibitem[Yang et al.(2013)]{yang13} Yang, W. M., Bi, S. L., Meng, X. C., Liu, Z. 2013, \apj, 776, 112
\bibitem[Yang et al.(2011)]{yang11} Yang, W. M., Meng, X. C., Bi, S. L., et al. 2011, \apj, 731, L37
\bibitem[Yang \& Tian(2017)]{yang17} Yang, W., \& Tian, Z. 2017, ApJ, 836, 102
\bibitem[Zhang et al.(2004)]{zhan04} Zhang, F., Han, Z., Li, L., Hurley, J. R. 2004, A\&A, 414, 117
\end{thebibliography}
\end{document}